\input harvmac
\input tables.tex

\def\fund{  \> {\vcenter  {\vbox  
              {\hrule height.6pt
               \hbox {\vrule width.6pt  height5pt  
                      \kern5pt 
                      \vrule width.6pt  height5pt }
               \hrule height.6pt}
                         }
                   }
           \>\> }

\def\antifund{  \> \overline{ {\vcenter  {\vbox  
              {\hrule height.6pt
               \hbox {\vrule width.6pt  height5pt  
                      \kern5pt 
                      \vrule width.6pt  height5pt }
               \hrule height.6pt}
                         }
                   } }
           \>\> }

\def\sym{  \> {\vcenter  {\vbox  
              {\hrule height.6pt
               \hbox {\vrule width.6pt  height5pt  
                      \kern5pt 
                      \vrule width.6pt  height5pt 
                      \kern5pt
                      \vrule width.6pt height5pt}
               \hrule height.6pt}
                         }
              }
           \>\> }

\def\symbar{  \> \overline{ {\vcenter  {\vbox  
              {\hrule height.6pt
               \hbox {\vrule width.6pt  height5pt  
                      \kern5pt 
                      \vrule width.6pt  height5pt 
                      \kern5pt
                      \vrule width.6pt height5pt}
               \hrule height.6pt}
                         }
              }
           } \>\> }


\def\atev{${\bf 8_v}$}
\def\ates{${\bf 8_s}$}
\def\atec{${\bf 8_c}$}

\def\ntwo{${\cal N}=2$}
\def\none{${\cal N}=1$}

\def\rarr{\rightarrow}

\def\NO{Nielsen-Olesen}

\def\npb#1#2#3{{Nucl.~Phys.} {\bf B#1} (#2) #3}
\def\prd#1#2#3{{Phys.~Rev.} {\bf D#1} (#2) #3}
\def\plb#1#2#3{{Phys.~Lett.} {\bf #1B} (#2) #3}

\def\hth{hep-th}

\lref\NAD{
N. Seiberg, {\it Electric-Magnetic Duality in Supersymmetric
Non-Abelian Gauge Theories}, \npb{435}{1995}{129}, \hth/9411149.}

\lref\kinsone{K. Intriligator and N. Seiberg,
{\it Phases of N=1 supersymmetric gauge theories in four dimensions.},
\npb{431}{1994}{551}, hep-th/9408155.}

\lref\kinstwo{K. Intriligator and N. Seiberg,
{\it Duality, Monopoles, Dyons, Confinement and Oblique Confinement in 
Supersymmetric $SO(N_c)$ Gauge Theories},
\npb{444}{1995}{125}, \hth/9503179.}

\lref\kinsrev{K. Intriligator and N. Seiberg,
{\it Lectures on Supersymmetric Gauge Theories and Electric-Magnetic
 Duality,} Nucl.~Phys.~Proc.~Suppl. {\bf 45BC} (1996) 1,
 \hth/9509066.}

\lref\nsewone{N. Seiberg, E. Witten, {\it Monopole Condensation And Confinement
In $N=2$ Supersymmetric Yang-Mills Theory}, Nucl. Phys. B426 (1994) 19,
hep-th/9407087.}

\lref\nsewtwo{N. Seiberg, E. Witten, {\it Monopoles, Duality and Chiral 
Symmetry Breaking in N=2 Supersymmetric QCD}, Nucl. Phys. B431 (1994)
484, hep-th/9408099.}

\lref\ppspin{P. Pouliot, 
{\it Chiral Duals of Non-Chiral SUSY Gauge Theories},
\plb{359}{1995}{108}, \hth/9507018.}

\lref\ppmsone{P. Pouliot and M.J. Strassler,
{\it A Chiral $SU(N)$ Gauge Theory and its Non-Chiral $Spin(8)$ Dual},
\plb{370}{1996}{76}, \hth/9510228.}

\lref\ppmstwo{P. Pouliot and M.J. Strassler,
{\it Duality and Dynamical Supersymmetry Breaking in $Spin(10)$ with a
Spinor}, \plb{375}{1996}{175}, \hth/9602031.}

\lref\mltispnr{M. Berkooz, P. Cho, P. Kraus and M.J. Strassler, {\it
Dual descriptions of SO(10) SUSY gauge theories with arbitrary numbers
of spinors and vectors}, \hth/9705003. To be published in Phys.Rev.D.}

\lref\LRW{E.J. Weinberg, D. London and J.L. Rosner,
{\it Magnetic monopoles with Z(N) charges}, \npb{236}{1984}{90}.}

\lref\NielOle{H.B. Nielsen, P. Olesen, {\it Vortex line models for 
dual strings}, Nucl. Phys. B61 (1973) 45. (Reprinted in *Rebbi,
C. (ed.), Soliani, G. (ed.): Solitons and Particles.)}

\lref\exactmono{F.A. Bais and R. Laterveer, {\it Exact Z(N) monopole
solutions in gauge theories with nonadjoint Higgs representations},
\npb{307}{1988}{487}.}

\lref\witten{E.Witten, {\it Current Algebra, Baryons, and Quark Confinement},
\npb{223}{1983}{433}. (Reprinted in *Rebbi,
C. (ed.), Soliani, G. (ed.): Solitons and Particles.)}
\lref\MQCD{E. Witten, {\it Branes and the dynamics of QCD}, hep-th/9706109;
A. Hanany, M.J. Strassler and A. Zaffaroni, {\it Confinement
and strings in MQCD}, hep-th/9707244}
\lref\chiralM{J. Lykken, E. Poppitz and S. P. Trivedi, {\it
Chiral gauge theories from D-branes.}  hep-th/9708134.}

\lref\accid{R.G. Leigh and M.J. Strassler,
{\it Accidental symmetries and N=1 duality in supersymmetric gauge
theory}, \npb{496}{1997}{132}, hep-th/9611020; J. Distler and
A. Karch, {\it N=1 dualities for exceptional gauge groups and quantum
global symmetries},  hep-th/9611088.}

\lref\seibone{
N. Seiberg, 
{\it Exact Results on the Space of Vacua of 
Four Dimensional SUSY Gauge Theories},
\prd{49}{1994}{6857}, \hth/9402044.}

\Title{\vbox{\rightline{hep-th/9709081}
\rightline{IASSNS--HEP--97/98}}}
{\vbox{\centerline{Duality, Phases, Spinors and Monopoles} \centerline{}
\centerline{in SO(N)
and Spin(N) Gauge Theories}}}

\centerline{ Matthew J.~Strassler }
\smallskip{\it
\centerline{School of Natural Sciences}
\centerline{Institute for Advanced Studies}
\centerline{Princeton, NJ 08540, USA}}
\centerline{ \tt strasslr@ias.edu}

\vskip .2in

\vglue .3cm

\noindent

Four-dimensional N=1 supersymmetric $Spin(N)$ gauge theories with
matter in the vector and spinor representations are considered.  Dual
descriptions are known for some of these theories.  It is noted that
when masses are given to all fields in the spinor representation, the
dual gauge group $G$ breaks to a group $H$ such that $\pi_2(G/H)=Z_2$.
The quantum numbers of the associated $Z_2$ monopole and those of the
massive spinors are shown to agree, suggesting that the monopole is
the image of the massive spinors under duality.  It follows that
electric sources in the spinor representation, needed as test charges
to determine the phase of an $SO(N)$ gauge theory, can be introduced as
$Z_2$-valued magnetic sources in the dual nonabelian gauge theory.
This fact is used to study the phases of $SO(N)$ gauge theories with
matter in the vector representation.

\Date{9/97}

\newsec{Overview}

Duality in field theory and string theory has become a central area of
research in recent years.  Although a great deal has been learned
at  both the technical and the conceptual level, there are still many
remaining questions. In this paper I examine some outstanding
issues involving the physics of \none\ supersymmetric gauge theories
in four dimensions.

After the original Seiberg-Witten solutions to \ntwo\ gauge theory
were found \refs{\nsewone,\nsewtwo}, duality was discovered in \none\
nonabelian supersymmetric gauge theories by Seiberg \NAD.  In \NAD\
and \kinstwo, the physics of duality for $SO(N)$ gauge
theories\foot{The distinction between $SO(N)$ and $Spin(N)$, its
double cover, is essential for the topological arguments used in this
paper.} with $N_f$ fields in the vector representation was studied.
The $SO(N)$ gauge theories were shown to have magnetic and sometimes
dyonic dual descriptions.  The different phases of these theories were
described.  For $N_f\geq 3(N-2)$ the theories are not asymptotically
free and are described as being in the ``free electric phase''.  In
this case their dual descriptions are strongly coupled.  For
$3(N-2)>N_f>{3\over 2}(N-2)$, the theory moves into the ``non-abelian
Coulomb phase'' where all descriptions are strongly coupled and the
low energy physics is that of a non-trivial conformal field theory.
At still lower $N_f$ the theory becomes very strongly coupled, while
one of its dual descriptions loses asymptotic freedom and becomes a
good weakly coupled description in the infrared.  This description has
an $SO(N_f-N+4)$ gauge group. If the dual gauge group is unbroken,
this phase is called the ``free magnetic phase'' since magnetic
degrees of freedom are free at long distances.  If the dual
description is an abelian gauge theory ($N_f=N-2$) then the original
description is in the ``Coulomb phase'', a special case of the free
magnetic phase.  For $N_f=N-3$ the theory enters the ``confining
phase''. The weakly coupled dual gauge theory is completely broken by
the Higgs mechanism; from the point of view of the original
description, the fields which have condensed are magnetically charged,
leading to confinement of the original fields.  Similar behavior
persists for $N_f=N-4$. For lower values of $N_f$ (except $N_f=0$) the
theory has no vacuum.

The arguments of \refs{\NAD,\kinstwo} were based on a interwoven
assemblage of powerful circumstantial evidence.  It would be nice,
however, to strengthen the arguments further, particularly in the
discussion of the various phase structures and their properties.  To
this end, it would be especially useful to be able to introduce
sources which are in the spinor representation of the gauge group.
Such sources have charges which cannot be screened by massless fields
of the $SO(N)$ theory, and so a Wilson loop in this representation
should be a good diagnostic for the phase of the theory.  How can the
Wilson loop in the spinor representation be introduced into the dual
description of the theory?  How is the electrically charged source
mapped under duality?  If we study $SO(2)$ without matter, then the
answer is known; as a consequence of the usual electric-magnetic
duality of the free classical Maxwell equations, the electrically
charged source of the first description is simply a magnetically
charged source of the other.  But such a straightforward
transformation is not possible in the non-abelian case.  Certainly the
appropriate duality transformation cannot be visible in the classical
equations of the theory.

To resolve this question, I continue with my historical review.
Pouliot soon discovered that $Spin(7)$ with fields in the spinor
representation is dual to a chiral, $SU(N)$ gauge theory \ppspin.
Further generalizations of this theory then followed.  In \ppmsone,
Pouliot and the author showed that $Spin(8)$ with $N_f$ fields in the
\atev\ and one field in the \ates\ representation is dual to a chiral
$SU(N_f-4)$ gauge theory.\foot{This and the other duality
transformations used in this paper are summarized in the Appendix.}
Furthermore, when the \ates\ is given a mass, the dual $SU(N_f-4)$
theory is broken to $SO(N_f-4)$, which is a dual description of
$Spin(8)$ with $N_f$ fields in the \atev.

In this we see the answer to the question posed above.  When
$SU(N_f-4)$ breaks to $SO(N_f-4)$, a topologically stable monopole
carrying a $Z_2$ charge is found in the theory \LRW.  This is because
$\pi_2[SU(N)/SO(N)]=\pi_1[SO(N)] = Z_2$, for $N>2$.  As I will argue
below, the massive spinor of the $Spin(8)$ theory is mapped under
duality to the $Z_2$ monopole of the broken $SU(N_f-4)$ theory.
Consequently, the Wilson loop in the spinor representation of
$Spin(8)$ is mapped to the 't Hooft loop in the magnetic $Z_2$
representation of the low-energy $SO(N_f-4)$ gauge theory.  This is
natural, since all charges of the spinor, except for its $Z_2$ charge
under the $Z_2\times Z_2$ center of $Spin(8)$, will be screened by the
light fields in the vector and adjoint representations.

The work of \ppmsone\ was further generalized to $Spin(10)$ gauge
theories with a number of fields in the ${\bf 10}$ representation and
one \ppmstwo\ or more \mltispnr\ fields in the ${\bf 16}$
representation.  Although masses for ${\bf 16}$'s cannot be introduced
in $Spin(10)$, they can be added when the theory is broken to a
smaller $Spin$ group.  The dual theories always contain an $SU$
factor, which is broken to an $SO$ gauge theory if and only if all
spinors are massive.  Again the spinors appear as monopoles carrying a
$Z_2$ charge.  More details will be given in sections 2 and 3.\foot{
Although these theories certainly have states with both electric and
magnetic charge, I have chosen not to discuss dyons and dyonic
duals\refs{\kinstwo,\kinsrev} in this paper.}

An aside about notation: since the magnetic monopoles in question are
found in theories which are themselves ``magnetic'' descriptions of
theories that may be in the free magnetic phase, there is obvious room
for confusion.  I therefore will abandon the commonly used terminology
of ``electric theory'' and ``magnetic theory'', which are conventional
in any case.  Instead I will refer to the $Spin(N)$ model with the
massive field(s) in the spinor representation as the ``A theory'', and to
its dual as the ``B theory''.

Returning to the physics, one can use the insight described above to
study the phases and dualities of these models.  For example, if the
$Spin(8)$ theory with a massive spinor is in the free magnetic phase,
so that the low-energy $SO(N_f-4)$ gauge group of the B theory is
weakly coupled in the infrared, then semiclassical physics of the B
theory monopoles implies that the spinors of the A theory are
unconfined and have a $(\log r)/r$ potential between them.  If some
additional fields in the A theory are given masses, so that the A
theory enters the confining phase, then the B theory gauge group is
completely broken via the Higgs mechanism, a \NO\ string soliton
\NielOle\ is present, and the spinors/monopoles are explicitly
confined by a linear potential. A more detailed discussion of phases,
supporting previous results of \refs{\NAD,\kinstwo,\kinsrev}, will be
given in section 4.

It is amusing that the $SU\rarr SO$ breaking pattern and the
topological relation $\pi_2[SU(n)/SO(n)] = Z_2$ have appeared before
in the context of strongly coupled $SO(N)$ gauge theories,
specifically in Witten's work on current algebra \witten. It is
perhaps important to emphasize the differences, to avoid confusion.
In Witten's conjecture, the $SU(n)$ and $SO(n)$ groups in the coset
are flavor symmetries of matter fields, with $n=N_f$, and with the
coset being the pion moduli space resulting from chiral symmetry
breaking. This breaking leads to a finite tension global string
soliton -- a two-dimensional Skyrmion -- which Witten suggested was
the string responsible for confinement of spinors.  Here the situation
is very different.  The $SU(n)$ and $SO(n)$ groups are gauge groups of
a dual description; all physical states are invariant under them.  The
coset is due not to chiral symmetry breaking, which does not occur in
these gauge theories, but to spontaneous gauge symmetry
breaking driven by a parameter in the Lagrangian or an expectation
value for a gauge singlet field.  The topology of gauge breaking leads
to an {\it unconfined} monopole in the B theory.  Confinement only
occurs if the $SO(n)$ group is itself completely broken.  The string
which is responsible for confinement of spinor particles is a \NO\
string of the dual theory, not a Skyrmion string built on the space of
gauge invariant vacua.

\newsec{General Expectations}

Let us first consider what we might expect the physics to look like.
If the A theory is strongly coupled, the fields of the A theory should
not, in general, be visible in the B theory.  The quarks of QCD cannot
be seen in the physics of the chiral Lagrangian; electrons
cannot be seen in the Landau-Ginsburg theory of
superconductivity.  Massless or light states should be especially
difficult to find, as they are deeply involved in the dynamics of the
theory.  Even massive particles may not be visible if all of their
charges can be screened.  For example, if both massive and massless
spinor representations of $Spin(N)$ are present, a massive spinor will
generate a massless spinor cloud around it, even if the dynamics are
relatively weakly coupled.  These neutral bound states of the A theory
will be visible as gauge singlets of the B theory, but the original
spinor fields themselves will not be visible.

By contrast, if the A theory has massive spinors but no massless ones,
then the fact that spinor representations have a charge under the
center of $Spin(N)$ becomes important.  The light fields of the theory
are all representations of $SO(N)$ and so are neutral under the group
$Z_2=Spin(N)/SO(N)$.  They therefore cannot screen the $Z_2$ charge of
the massive spinors.  Although a given spinor can surround itself with
a cloud of light fields, its $Z_2$ charge cannot be removed.  On the
other hand, a state containing two spinors need not be visible (if the
spinors are not widely separated) since the two as a pair may be
screened by the light fields.

One may then expect that the lightest state with non-zero $Z_2$ charge
might be visible in the B theory.  Since this state must disappear
both when the spinor mass is taken to infinity and when it is taken to
zero, it is natural to expect the masses of the B theory state and
that of the A theory spinor are correlated.  But there is no reason to
expect the masses to be related in a simple way; the binding energy
between the spinor and its surrounding cloud may be very large.

What if the A theory has many spinors?  If all of the spinors have the
same mass, then the lightest states in the A theory with non-zero
$Z_2$ charge have a certain degeneracy.  It is natural to expect
this degeneracy to be visible in the B theory as well; the flavor
symmetries of the lightest spinors ought to be visible in the B
theory.

In conclusion, the B theory may be expected to contain heavy states
with a $Z_2$ charge, whose mass is correlated with the spinor
mass(es), and whose flavor symmetries match those of the spinors.

It is natural that the state carrying this $Z_2$ charge be a magnetic
monopole, given the situation in \ntwo\ gauge theories \nsewtwo.
Consider $SO(3)$ with a massless triplet and $2N_f<8$ doublets of mass
$m$, coupled to the triplet in an \ntwo-symmetric fashion, as the A
theory.  Quantum mechanically the $SO(3)$ gauge group is broken to
$SO(2)$, under which the massive doublets have charge $\pm 1/2$.  A
Maxwell electric-magnetic duality transformation on the $SO(2)$
converts the description to that of the B theory.  The electrically
charged massive doublets become heavy Dirac magnetic monopoles, with
magnetic charge $\pm 1/2$, of the B theory.  Since the light fields of
the A theory, the triplet and the gauge bosons, have charge $0,\pm 1$,
the heavy doublets/monopoles carry a conserved $Z_2$ quantum number.
The situation considered in this paragraph will emerge as a special
case of the results given below.

\newsec{Flavor Representations}

The strongest evidence that the massive particle in the spinor
representation of the A theory appears in the B theory as a monopole
comes from the transformation properties of the spinor and monopole
under flavor groups of the two theories.  It will also become clear in
this section that monopoles only arise in the B theory when {\it all}
of the spinors in the A theory are massive, consistent with the discussion
of section 2.

\subsec{The massive spinor of $Spin(8)$}

Consider, as the A theory, $Spin(8)$ with $N_f$ fields $V^i$ in
the \atev\ and a single massive spinor $P$ in the \ates.  The
superpotential is $W=\half mPP$.  For bookkeeping purposes, let us
separate the $V$ fields into $V^i$, $i=1,\dots,N_f-k$, and $\hat V^r$,
$r=1,\dots,k$.  We may then go to the point in moduli space where
$\vev{\hat V^r\hat V^s} = v^2\delta^{rs}$.  In this vacuum the gauge
symmetry is broken to $Spin(8-k)$ and the $SU(N_f)$ flavor symmetry is
broken to $SU(N_f-k)\times SO(k)$.  Actually the last factor, which is
a diagonal subgroup of the original flavor and color groups, is
$Spin(k)$; the field $P$ becomes a (generally reducible)
eight-dimensional bispinor under $Spin(8-k)\times Spin(k)$.  For
example, if $k=2$, then $P$ is a ({\bf 4}, {\bf +})  $+$ ({\bf
$\bar 4$}, {\bf $-$}) of $Spin(6)\times Spin(2)$, while if $k=3$, then
$P$ is a $({\bf 4},{\bf 2})$ of $Spin(5)\times Spin(3)$.

As described in the Appendix, the B theory has a $SU(N_f-4)$ gauge
group, with matter $S$ in the symmetric representation and $Q_i,\hat
Q_r$ in the antifundamental representation.  It also has gauge
singlets $N^{ij},N^{ir},N^{rs}$ and $T$.  When the singlet $N^{rs}=V^rV^s$
has an expectation value $N^{rs}=v^2\delta^{rs}$, the effective
superpotential becomes
\eqn\ateBzro{ W =
\sum_{i,j=1}^{N_f-k}{1\over \mu_1^2}N^{ij}Q_iSQ_j +
\sum_{r=1}^k{v^2\over \mu_1^2}\hat Q_rS\hat Q_r +{1\over \mu_2^{N_f-5}}T\det S
+ mT } 
The $\mu_i$ are parameters of dimension one needed for
dimensional consistency (and other issues) \refs{\kinstwo,\kinsrev};
their presence and physical meaning will be irrelevant for this paper.
The singlets $N^{ir}$ are the images under duality of the
$Spin(8-k)$-singlet components of the $V^i$; they have decoupled from
this superpotential in the standard way with the help of a field
redefinition. The flavor group is clearly broken to $SU(N_f-k)\times
SO(k)$, in agreement with the A model.  The F-flatness condition
$\partial W/\partial T=0$ ensures that $\vev{\det S}$ is nonzero, and
so $SU(N_f-4)$ is broken to $SO(N_f-4)$, leading to a $Z_2$ monopole
\LRW.  As a consequence of their coupling to $S$, the $\hat Q_r$
become massive, and they each have a single zero mode in the presence
of the monopole.  These zero modes transform in the vector
representation of $SO(k)$, and therefore, after quantization, the
$Z_2$ monopole will transform as a spinor of $Spin(k)$.  This agrees
with the global charges of $P$.

\subsec{The massive spinor of $Spin(10)$}

Next consider, as the A model, $Spin(10)$ with $N_f$ fields $V^i$ in
the ${\bf 10}$ and a single spinor $P$ in the ${\bf 16}$.  Since the
spinor representation is chiral, we cannot write a mass term for $P$,
but mass terms can be written if $Spin(10)$ is broken to a smaller
$Spin$ group; in $Spin(10)$ language, $P$ may become massive by
coupling to a vector which acquires an expectation value.  Again let
us separate the fields into $V^i$, $i=1,\dots,N_f-k$, and $\hat V^r$,
$r=1,\dots,k$.  We will take the superpotential to be $W = y\hat V^1
PP$.  At the point in moduli space where $\vev{\hat V^r\hat V^s} =
v^2\delta^{rs}$, the spinor $P$ has mass $yv$, the gauge symmetry is
broken to $Spin(10-k)$ and the $SU(N_f)$ flavor symmetry is broken to
$SU(N_f-k)\times Spin(k-1)$.  For very small $y$ the last factor is
$Spin(k)$, and we may take the limit of large $v$ and small $y$,
holding $yv$ fixed, for the purposes of discussing quantum numbers.
The field $P$ decomposes into a generally reducible sixteen-component
bispinor representation of $Spin(10-k)\times Spin(k)$.  For example,
if $k=2$, then $p$ is a ({\bf \ates}, {\bf +}) $+$ ({\bf \atec}, {\bf
$-$}) of $Spin(8)\times Spin(2)$, while if $k=5$, then $p$ is a $({\bf
4},{\bf 4})$ of $Spin(5)\times Spin(5)$.

As reviewed in the Appendix, the B theory has $SU(N_f-5)$ gauge group
with matter $S$ in the symmetric tensor representation, $Q_i,\hat Q_r$
in the antifundamental representation, and $F$ in the fundamental
representation, along with gauge singlets $N^{ij},N^{ir},N^{rs}$ and
$Y^i,\hat Y^r$.  When the singlet $N^{rs}$ has an expectation value
$N^{rs}=v^2\delta^{rs}$, the effective superpotential becomes
\eqn\tenBzro{ W = {\det S\over \mu_2^{N_f-8}} +
\sum_{i,j=1}^{N_f-k}{1\over \mu_1^2}N^{ij}Q_iSQ_j +
\sum_{r=1}^k{v^2\over \mu_1^2}\hat Q_rS\hat Q_r +
{1\over\mu_3^2}(\sum_{i=1}^{N_f-k}Y^iQ_i + \sum_{r=1}^k\hat Y^r\hat
Q_r) F + y\hat Y^1 } 
The flavor group is broken to $SU(N_f-k)\times SO(k-1)$, with the
latter factor extended to $SO(k)$ for small $y$, in agreement with the
A model.  The F-flatnesss condition $\partial W/\partial Y^1=0$
equation for $Y^1$ ensures that $\vev{\hat Q_1 F}$ is
non-zero, breaking $SU(N_f-5)$ to $SU(N_f-6)$.  Under this breaking
$S$ decomposes into a symmetric tensor $s$, a fundamental $f$,
and a singlet $z$.  Through the condition  $\partial W/\partial z=0$
he expectation value for $\hat Q_1$  then requires $\det
s \propto  \hat Q_1\hat Q_1\propto y\mu_3^2$.  This breaks the theory
to $SO(N_f-6)$, leading to a $Z_2$ monopole \LRW.  As a consequence of their
coupling to $S$, the $\hat Q_r$ become massive when $\det S$ gets an
expectation value, and they each have a single zero mode in the
presence of the monopole.  Since these zero modes transform in the
vector representation of $SO(k)$, the $Z_2$ monopole will transform as
a spinor of $Spin(k)$. As before, this agrees with the global charges
of $P$.

As an additional check, let us examine what happens if we take $k=2$,
and let the fields $\hat V^r$ have expectation values such that
$\vev{V^1V^1}=\vev{V^2V^2}=0$, $\vev{V^1V^2}=v^2$.  The A theory is
broken to $Spin(8)$, with $N_f-2$ vectors $V^i$ in the \atev, and with
the spinor $P$ decomposing into a field $P_s$ in the \ates\ and a
field $P_c$ in the \atec.  The $Spin(10)$ superpotential we take to be
\eqn\PsPcAspot{
W = [y_s \hat V^1 +y_c\hat V^2]PP } 
One can check that in the low-energy $Spin(8)$ theory, the mass of
$P_s$ is proportional to $y_sv$ and that of $P_c$ is proportional to
$y_cv$.  The flavor symmetry of the model is $SU(N_f-2)\times Spin(2)$
if $y_s=y_c$, with the $Spin(2)$ broken otherwise; $P_s$ and $P_c$
have opposite charge under the $Spin(2)$.  It is interesting to
consider taking one mass to zero holding the other fixed, or
alternatively holding the first fixed and taking the other to
infinity.  What happens, in these two limits, in the B theory?

The B theory has superpotential
\eqn\PsPcBspot{\eqalign{
W = {\det S \over \mu_2^{N_f-8}}+&
\sum_{i,j=1}^{N_f-2}{1\over \mu_1^2}N^{ij}Q_iSQ_j +
{v^2\over \mu_1^2}\hat Q_1 S\hat Q_2 +
{1\over\mu_3^2}(\sum_{i,j=1}^{N_f-2}Y^iQ_i + \hat Y^1\hat Q_1+\hat
Y^2\hat Q_2) F \cr
& + y_s\hat Y^1+ y_c\hat Y^2\cr}}
The F-term equations for $\hat Y^1, \hat Y^2$ imply that $\hat
Q_1,\hat Q_2,F$ get expectation values, which by the D-term equations
must all lie along the same direction in the color group, breaking the
gauge group to $SU(N_f-6)$.  The field $S$ decomposes as above, and
the condition $\partial W/\partial z=0$ then requires $\det s\propto
Q_1Q_2\propto y_sy_c\mu_3^2/\sqrt{|y_s|^2+|y_c|^2}$.  This breaks the
gauge group to $SO(N_f-6)$ and generates a $Z_2$ monopole solution.
The flavor group includes $SU(N_f-2)$, with an additional $SO(2)$
factor if $y_s=y_c$.  Thus, only when $y_s=y_c$ must the stable
monopole be a doublet of $SO(2)$, according with the A theory.  
Furthermore, if
either one is massless, it can screen the other, making both
invisible.  We can see these effects in the limits $y_sy_c=0$ or
$\infty$.  As we take $y_s$ or $y_c$ to zero, holding the other fixed,
so that the $SU(N_f-6)\rarr SO(N_f-6)$ breaking scale becomes very
low, the monopole becomes light, disappearing in the $y_sy_c=0$ limit
where the $SU(N_f-6)$ symmetry is restored.  In the limit $y_s$ or
$y_c$ goes to infinity, with the other held fixed, the $SU(N_f-6)\rarr
SO(N_f-6)$ breaking scale goes to a constant, keeping the monopole at
a finite mass.

One may perform a similar analysis by breaking the $Spin(10)$
theory with one spinor to
$Spin(4)$, leaving a $Spin(6)$ flavor group. The spinors transform in the
$({\bf 2,4})$ and $({\bf \bar 2,\bar 4})$ representations of
the $Spin(4)\times Spin(6)$ group.  If half the spinors are much more
massive than the others, the flavor group is reduced to $Spin(5)$,
with the light spinors transforming in the $({\bf 2,4})$ of
the $Spin(4)\times Spin(5)$ group.  It can be easily
checked that in the process the $SO(6)$ flavor group
of the B theory is reduced to $SO(5)$, leaving the monopole
with the correct quantum numbers.

\subsec{Multiple Spinors}

Finally, consider $Spin(10)$ with $N_P$ fields $P_a$ in the ${\bf
16}$ representation and $N_f$ fields $V^i$ in the ${\bf 10}$
representation \mltispnr.  Breaking the theory to $Spin(9)$ or below,
one may add masses for some or all of these fields.  However, the B theory
is remarkably complicated, and a complete analysis is difficult to
perform, although certain simple observations are possible.  It can be
shown that if all $N_P$ spinors are given a mass, then the
$SU(N_f+2N_P-7)\times Sp(2N_P-2)$ gauge symmetry of the B theory is
broken to $SO(N_f-6)$, a breaking pattern which predicts a $Z_2$
monopole.  If any of the spinors are massless, the breaking to an $SO$
group does not take place and there is no monopole in the B theory.
Furthermore, since the appearance of the $SU(N_f)$ global symmetry in
the B theory is quite simple, and is in fact identical to that of the
case $N_P=1$ discussed earlier, we can identify the quantum numbers of
the monopole under this symmetry.  In particular, if the $Spin(10)$
theory is broken to $Spin(10-k)$ by expectation values for $k$ of the
fields $V^i$, then the $SU(N_f)$ global symmetry is broken to
$SU(N_f-k)\times Spin(k)$, with all the spinors $P_a$ transforming as
bispinors under $Spin(10-k)\times Spin(k)$.  The effect on the B
theory is as before; $k$ of the $N_f$ fields $Q_i$ in the
antifundamental representation of $SU(N_f+2N_P-7)$ develop couplings
$Q_i S Q_i$ to the symmetric tensor field $S$.  The flavor symmetry is
broken thereby to $SU(N_f-k)\times SO(k)$.  When the field $S$
acquires an expectation value and leads to a monopole solution, the
$k$ distinguished $Q_i$ develop zero modes in the presence of the
monopole, making it a spinor of the global symmetry $SO(k)$, as
expected.

The monopole will also transform under the other flavor symmetry of
the theory, the one which rotates the $N_P$ spinors into each other.
This global symmetry, which is $SU(N_P)$ in the A theory, is reduced
to $SU(2)$ in the B theory \mltispnr, with the full $SU(N_P)$ only being
realized quantum mechanically in the B theory, as a quantum accidental
symmetry \accid.  The $N_P$ spinors $P$ transform as an $N_P$
dimensional representation (a symmetric combination of $N_P-1$
doublets) of this $SU(2)$, a remarkable structure not previously seen
in duality.  Consequently, a prediction of the spinor-monopole
identification is that the B model should have zero modes which make
the monopole an $(N_P-1)$-index multispinor under the $SU(2)$ flavor symmetry.
It would be a remarkable check on the results of this paper if this
pattern of zero modes could be confirmed.  To verify it, however,
requires an analysis of the complex and intricate breaking pattern in
the B model -- a challenge for the reader.

\subsec{Summary}

I have shown in several examples that, where global $Spin(k)$ flavor
symmetries arise as a result of symmetry breaking in the A theory, the
monopoles of the B theory and the massive spinors of the A theory
transform in the same way under them. Furthermore, the monopoles only
exist in those theories in which {\it all} spinors are massive; the
presence of even one massless spinor in the A theory eliminates the
$SU\rarr SO$ breaking pattern in the B theory.


\newsec{The Physics of Various Phases}

Having established the plausibility of the spinor-monopole
identification in these theories, I now turn to a study of their
phases.  My conclusions will support those of previous authors
\refs{\seibone,\nsewone,\nsewtwo,\kinsone,\NAD,\kinstwo,\kinsrev}.

I will focus my discussion on the $Spin(8)$ theory and its
dual description.  (To repeat the analysis for the $Spin(8-k)$ and
$Spin(10-k)$ theories considered in the previous section requires
minor, inessential modifications.)  Throughout this section I will
always be referring to the phase of the {\it low-energy} A theory,
that is, the theory of $Spin(8)$ with a certain number of {\it
massless} fields in the vector representation, except when explicitly
noted.

\subsec{The Coulomb Phase}

Let us consider, as the A theory, $Spin(8)$ with one spinor $P$ of
mass $m$, $M_f$ vectors $V^r$ of mass $m_r$, and six massless vectors
$v^i$.  This is known as the Coulomb phase of the A theory, since at
the generic point in moduli space the six massless vectors break
$Spin(8)$ to pure $Spin(2)$ without matter.  For non-zero $m,m_r$, the
B theory, with gauge group $SU(M_f+2)$, will be broken to $SO(2)$.

Before studying this theory, it is worth reviewing the results
expected when the spinor mass $m$ is infinite but the vector masses
$m_r$ are finite \refs{\NAD,\kinstwo}.  If $m_r=0$ the B theory is an
$SO(M_f+2)$ gauge theory with $M_f+6$ fields $Q_r,q_i$ in the vector
representation.  Let us give expectation values to the six fields
$v^i$ and masses to the $M_f$ fields $V^r$.  The A theory gauge group
becomes $SO(2)$, with $M_f$ massive fields $V^r$ and twenty-seven
massive gauge bosons of charge $0,\pm 1$, and some 't Hooft-Polyakov
magnetic monopoles.  In the B theory, the six fields $q_i$ become
massive while the other matter fields develop expectation values
$\vev{Q^rQ^r}\neq 0$, $r=1,\dots,M_f$.  This breaks the B theory gauge
group to $SO(2)$, with six massive fields $q^i$ and
$[(M_f+2)(M_f+1)/2]-1$ massive gauge bosons of charge $0,\pm 1$, and
some 't Hooft-Polyakov monopoles.  In both descriptions, these
monopoles carry magnetic charge $\pm 1$ in units where the minimum
Dirac value is $\pm 1/2$.  (This magnetic charge is additive, in
contrast to the $Z_2$ monopoles discussed in section 2 and 3 where the
charge was only defined mod 2.)  From the duality of \kinstwo, it can
be seen that if the $V^r$ have equal masses and the $v^i$ have equal
vacuum expectation values, then the number of monopoles in the A
theory is at least six and the number of monopoles of the B theory is
at least $M_f$.

In short, the $V^r$ are visible as monopoles in the B theory while the
$q_i$ are visible as monopoles in the A theory. (The extent to which
the gauge bosons of the two theories are visible as monopoles has not
been fully explored.)
The reason that these states are visible, rather than screened, is
that there are no massless charged states at the generic point in
moduli space.  Quantum mechanically, it remains true throughout the
entire moduli space that no states electrically charged under the A
theory become massless.  This is analogous to the physics in $SO(3)$
with a single triplet, also known as pure \ntwo\ $SU(2)$ gauge theory
\nsewone.  The only particles which become massless are monopoles and
dyons of the A theory.  The absence of screening is therefore manifest
on the entire Coulomb branch, including at the origin of moduli space
where $Spin(8)$ is classically unbroken, and so all electric charges
of the A theory become identifiable magnetic charges in the $SO(2)$ B
theory.

Let us now consider finite $m$, with $m\gg m_r$, and repeat this
analysis.  Since spinors carry electric charges $\pm 1/2$ when the A
theory is broken from $Spin(8)$ to $SO(2)$, it is natural to expect
that when the B theory is broken from $SU(M_f+2)$ to $SO(2)$, the
spinors will appear as monopoles of magnetic charge $\pm 1/2$.  To see
that the monopoles from the $SU(M_f+2)\rarr SO(M_f+2)$ breaking have
half the charge of those from the $SO(M_f+2)\rarr SO(2)$ breaking is
straightforward.  As in the original 't Hooft-Polyakov monopole, both
of these monopole solutions involve the winding, in some $SU(2)$
subgroup of the full B theory gauge group, of a field which is a
triplet under that $SU(2)$ \exactmono.  In the first case, the triplet
is a part of the field $S$, whose electric charges under the unbroken
$SO(2)$ are $\pm 2, 0$.  By contrast, the triplet involved in the
second monopole is a part of a field $Q_i$, whose charges are $\pm
1,0$ under the unbroken $SO(2)$.  Thus the two $SU(2)$'s are
normalized differently, and it follows that the magnetic charge of the
second monopole is twice that of the first.  Corroboration is provided
by the proof in \LRW\ that two $SU(M_f+2)/SO(M_f+2)$ monopoles either
can be completely unwound or can be deformed into an $SO(M_f+2)/SO(2)$
monopole.\foot{Whether the magnetic charge of a given
$SU(N_f-4)/SO(N_f-4)$ monopole under $SO(2)$ is positive or negative
depends on its orientation inside the $SO(N_f-4)$ group \LRW.}

In the case $m\gg m_r$ under discussion, where the breaking pattern
in the B theory is $SU(M_f+2)\rarr SO(M_f+2)\rarr SO(2)$, the
monopoles of half-integer charge are much heavier than those of
integer charge.  This nicely reflects the relative masses of the
particles in the A theory with half-integer and integer electric
charge.  But if $m\ll m_r$, then the spinors are light, the lightest
state of charge $\pm 1$ will not be a $V^r$ particle, and we should
not see stable heavy monopoles of charge $\pm 1$.  This is visible in
the B theory, where the expectation values for $Q_iSQ_i$ break the
gauge group first to $SU(2)$, generating no monopoles, and then the
expectation value for $\det S$ breaks it to $SO(2)$, generating
monopoles carrying an additive charge.  These monopoles carry magnetic
charge $\pm 1/2$, since they are built by winding the field $S$, and
so correspond to spinors.  The states of charge $\pm 1$ are light and
remain so as $m_r\rarr\infty.$

Thus, the various limits are consistent with expectations.  If we
take $m$ to infinity, $m_r$ fixed, we find that the monopoles with
half-integer charge become infinitely massive while those of integer
charge survive.  If we hold $m$ fixed and take $m_r$ to zero, the
monopoles with integer charges become light and disappear, while the
monopoles with half-integer charge remain as $Z_2$ monopoles of the
non-abelian B theory.  If we take $m_r$ to infinity and hold $m$
fixed, states with any magnetic charge survive in the B theory with finite
mass.  And if we hold $m_r$ fixed and take
$m$ to zero, all magnetic monopoles disappear from the B theory,
reflecting the expectation that all massive
$V^r$ will be screened by the massless $P$ field.

\subsec{The Confining Phase}

Let us consider the same theory as above, with one modification; let
us add a mass $\hat m_6$ for $v^6$.  Now, at the generic point in
moduli space, there is an unbroken pure $Spin(3)$ gauge group.  We
expect such a theory to confine throughout the moduli space, and will
focus on the point at the origin.

Of course, the test for confinement in $SO(N)$ gauge theories with
massless matter in the vector representation is generally to introduce
sources charged under the spinor representation of the group.  If
these sources have a potential energy linear in their separation ---
or equivalently, if a Wilson loop in the spinor representation has an
area law --- then the theory is confining.  Since the spinor is a
magnetic monopole in the B theory, the Wilson loop in the spinor
representation of the A theory is directly mapped to a 't Hooft loop
in the $Z_2$ magnetic representation of the B theory.  We may now
confirm that this theory is confining.  In particular, the B theory,
which for $\hat m_6=0$ is an $SU(M_f+2)$ gauge theory broken to
$SO(2)$, will be completely broken for non-zero $\hat m^6$.  As in the
Abelian Higgs model, magnetic flux will be confined into \NO\ strings
\NielOle, resulting in a linear potential for the monopoles.  This is
consistent with the conventional expectations based on the dual
Meissner effect, and agrees with \refs{\NAD,\kinstwo}.  The massive
fields in the vector representation also show confinement at
sufficiently short distances.

As before I consider several possible breaking patterns for the B
theory, in order to illustrate different aspects of the physics.  It
is useful to keep in mind the result from the previous section, that
both $V^r$ and $P$ are visible in the B theory if $m\gg m_r$, while
only $P$ is visible if $m\ll m_r$.

If $m\gg m_r\gg \hat m_6$, then both the massive $P$ and $V^r$
particles, previously visible as monopoles in the B theory, should be
confined.  A string with flux $1/2$ should break via pair production
of $P$ particles which are very heavy, while an assemblage of strings
with integer flux should break at much lower scales.  Consequently the
confinement of the $V^r$ should break down at distances short compared
with the scale at which confinement of $P$ particles is lost.  

Indeed this is seen in the B theory, which breaks as $SU(M_f+2)\rarr
SO(M_f+2)\rarr SO(2) \rarr 1$.  Our previous discussion indicated that
there are heavy and light monopoles of half-integral and integral
magnetic charge.  The breaking of $SO(2)$ by a field $Q_6$ of integral
electric charge confines magnetic flux in a \NO\ string configuration,
which carries an additive $Z$ charge.  The flux carried by the string
corresponds to magnetic charge $1/2$ under the broken $SO(2)$.  The
$SU(M_f+2)/SO(M_f+2)$ monopoles are therefore confined by a string of
flux $1/2$, while those of the $SO(M_f+2)/SO(2)$ theory are confined
by either two strings with flux $1/2$ or a single string with flux
$1$, the actual configuration being determined dynamically.
 When two monopoles of integral charge are pulled far
apart, and the total energy becomes very large, the broken $SO(2)$
theory will become sensitive to the fact that it is actually a broken
$SO(M_f+2)$ theory, and since $\pi_1[SO(M_f+2)] = Z_2$, the string
configuration with total flux $1$ will be able to unwind.  This
represents the screening of the $V^r$.  Only at much higher energies
-- much longer separations -- will the fact that $\pi_1[SU(M_f+2)] =
1$ become important and lead to breaking of the string of flux $1/2$.
When the spinor mass $m$ is taken arbitrarily large, the
$SU(M_f+2)\rarr SO(M_f+2)$ breaking scale goes to infinity, the $Z_2$
monopoles become infinitely massive, and the string of flux $1/2$ does
not break.

What happens if we increase $\hat m_6$ until $m\gg m_r\sim \hat m_6$?
One would expect that the confinement scale increases in energy to the
point that the massive $V^r$ particles cannot be far separated before
their strings break via pair production, and therefore that the $V^r$
particles and the strings of carrying flux 1 would not be found in the
B theory.  This expectation is fulfilled.  The B theory breaks
directly from $SU(M_f+2)\rarr SO(M_f+2)\rarr 1$ and its monopoles and
\NO\ strings carries only a $Z_2$ charge (since $\pi_1[SO(M_f+2)] =
Z_2$); the monopoles and strings associated to the $V^r$ have
disappeared.  This theory is an explicit and physically interesting
realization of the non-abelian generalization of the dual Meissner
effect.

By contrast, if $m_r\gg m\gg \hat m_6$ then the heavy $V^r$ particles can
be screened by pair production of $P$ particles.  The $SU(M_f+2)\rarr
SU(2)\rarr SO(2) \rarr 1$ breaking pattern in the B theory
correspondingly gives only light monopoles bound by \NO\ strings with
half-integral flux.  If we increase $\hat m_6$ so that $m_r\gg\hat
m_6\gg m$, then we expect the confinement scale in the A theory to be
so large that the strings can easily break via pair production of $P$
particles and should not be visible.  Indeed, in this limit the B
theory breaks directly from $SU(M_f+2)$ to nothing and has neither
monopoles nor strings.

In short, duality gives a picture of confinement essentially
consistent with conventional expectations, and provides a fully
non-abelian example of the dual Meissner effect.

\subsec{The Free Magnetic Phase}

If the identification of the $Z_2$ monopole with $P$ is accepted, it
can be used to confirm Seiberg's conception of the non-abelian free
magnetic phase \refs{\seibone,\kinsone,\NAD,\kinsrev}.  If the A theory is $Spin(8)$
with $N_f$ massless fields $V^i$ in the \atev\ and a massive spinor
$P$ in the \ates, then the B theory is an $SU(N_f-4)$ gauge theory
broken to $SO(N_f-4)$, with $N_f$ fields $Q_i$ in the $\bf{N_f-4}$
representation and some gauge singlets $N^{ij}$.  For $6\leq N_f<9$
the B theory has non-negative beta function, and so is weakly coupled
in the infrared.

For $N_f=6$, the theory is in the Coulomb phase discussed earlier; the
B theory has $SO(2)$ gauge symmetry.  Far out along the moduli space,
where the A theory is broken to $SO(2)$ also, ordinary Maxwell
electric-magnetic duality implies that electrically charged sources
carrying spinor charge in the A theory will appear as magnetically
charged sources in the B theory of charge $\pm
1/2$ \refs{\nsewtwo,\NAD,\kinstwo}. One may then argue that this
identification can be carried to the origin of moduli space, where
Maxwell duality cannot be directly used.  The spinor-monopole
identification lends further credence to this argument, since as shown
earlier the spinor indeed appears as a monopole of charge $\pm 1/2$.
It is known from $SO(N)$ duality that at the origin of moduli space,
the six fields $Q_i$ carrying electric charge under the B theory
become light, causing the gauge coupling of the B theory to run
logarithmically to zero at long distance.  Consequently, the coupling
governing interactions of magnetic sources goes logarithmically to
infinity.  We may conclude that the interaction between two static
spinor particles separated by a large distance $r$ takes the form $\log
r/r$, as argued in \kinsrev.\foot{Note that along the part of the moduli
space where $Spin(8)$ is broken to $Spin(3)$, the theory is related to
the \ntwo\ theories studied by Seiberg and
Witten \refs{\nsewone,\nsewtwo}; the discussion of this paragraph is of
course consistent with their results.}

For $N_f>6$ the situation has previously been less clear.  Far from
the origin of moduli space, the A theory is broken to $SO(2)$ with
massless matter. Because of the light charged fields, classical
Maxwell electric-magnetic duality cannot be used.  Interpolation to
the origin of moduli space is therefore nontrivial. However, the
arguments of this paper resolve any outstanding issues.  The B theory
has non-negative beta function, so at the origin of moduli space its
gauge coupling flows logarithmically to zero at long distance.  The
interaction energy between two static spinor sources in the A theory
is just that of two semiclassical $Z_2$ monopoles of the B theory; it
behaves as $\log r/r$ with a computable coefficient.

A comment about the difference between the confining phase
and the free magnetic phase should be made.  The presence of the
massless mesons $N^{ij}$, which are weakly coupled massless particles
of the B theory in the free magnetic, Coulomb and confining phases,
might lead at first glance to the misconception that electric charges
in the A theory are confined.  It is tempting to think of $N^{ij}$,
which carries the flavor quantum numbers of a bound state of two $V^i$
particles, as a confined system.  However, this image is clearly
inaccurate, as we have just seen. The presence of the massless mesons
is instead closely related to anomalies and chiral symmetries.
Similar analysis applies to the real-world pion system; it is well
understood that the presence of light pions does not in any way imply
confinement.

Perhaps this is also a good place to address the question of whether
it is reasonable in \none\ supersymmetry to trust the identification
of the massive monopole with the massive spinor.  Ideally, there would
be a limit in which the B theory was weakly coupled at the scale of
the monopole mass, so that one could show that the B theory
description of the A theory really did contain a monopole solution.
However, such a limit does not exist.  As I will now show, this
follows from the fact that the $Spin(8)$ theory with a massless spinor
does not have a free magnetic phase; {\it i.e.}, that the B theory
with unbroken $SU(N_f-4)$ gauge group always has a negative beta
function.\foot{The A theory has a free electric phase, and so
the reverse argument {\it does} work, as will be shown in the next
section.} After explaining the problem, I will give the strongest
argument that I can construct.

First, consider the following physical situation: at high energies the
A theory description is weakly coupled, and there is a spinor with
mass $m$ large compared with the scale $\Lambda_A$ where the A theory
coupling becomes strong.  The question is whether the B theory has a
monopole with the same properties as the A theory spinor.  Of course,
when we convert to B theory variables, they are strongly coupled at
the scale of order $m$ where $SU(N_f-4)$ breaks to $SO(N_f-4)$, as
they must be, since A theory perturbation theory works there.  Only
after the breaking takes place does the B theory begin to flow toward
weak coupling.  Thus, for large $m$, we cannot trust B theory
semiclassical arguments about soliton solutions.  It is then natural
to ask if by lowering the mass $m$ we can reach a regime where
semiclassical reasoning in the B theory will work and give a magnetic
monopole via the $SU(N_f-4)\rarr SO(N_f-4)$ breaking.  If this were
so, then we would expect that as $m$ is taken larger, the details of
the monopole solution would no longer be given by semiclassical B
theory physics, but the existence of at least one heavy, magnetically
charged state would be ensured.  Unfortunately, the B theory does not
become weakly coupled when $m$ is lowered, since the beta
function of the $SU(N_f-4)$ group is negative. As a result, for small $m$,
semiclassical physics fails completely; the spinor is too
light to be treated perturbatively in the A theory, while the B theory
remains strongly coupled at the $SU(N_f-4)\rarr SO(N_f-4)$ breaking
mass scale, making a semiclassical soliton analysis impossible.  In
fact, as $m\rarr 0$ the theory flows to a non-trivial conformal fixed point.

Nevertheless, an argument from a different point of view can still be
given, using the fact that the {\it low-energy} B theory is weakly
coupled.  Although semiclassical reasoning at short distances is
unreliable, topological reasoning at long distances is trustworthy.
Topology implies that the low-energy $SO(N_f-4)$ gauge theory can in
principle have a $Z_2$ monopole, since $\pi_1[SO(N_f-4)] = Z_2$
implies the consistency of a $Z_2$ Dirac string.  Independently of
duality, semiclassical arguments show that if the B theory is weakly
coupled at the scale where $SU(N_f-4)$ breaks to $SO(N_f-4)$, then a
soliton with long-range magnetic fields, whose total flux is
conserved, certainly exists.  If we then consider increasing the
coupling of the $SU(N_f-4)$ theory to a large value, as needed to
match onto the physics of the A theory, the details of the monopole,
such as its mass and core shape, will change in an uncontrollable way;
it may even decay to lighter states.  But despite these changes there
must still exist some state in the theory charged under a $Z_2$ and
surrounded by a long-range $SO(N_f-4)$ magnetic field.  This state
must be both heavy and small, since otherwise it will be in conflict
with the weakly coupled physics of $SO(N_f-4)$. For the arguments used
in this study of phases, merely the existence of such an object is
needed.

\subsec{The Free Electric Phase and Large $N_f$}

Again consider $Spin(8)$ with $N_f$ massless fields $V^i$ in the
\atev\ and a massive spinor $P$ in the \ates.  If $N_f\geq 18$, the
$Spin(8)$ theory loses asymptotic freedom.  The interesting physical
situation is then given by taking the B theory as weakly coupled
in the far ultraviolet; below its strong coupling scale $\Lambda_B$ it
flows to the weakly coupled A theory.  To make the B theory
renormalizable, we should consider it with superpotential $W=0$; the
resulting adjustment to the duality, given in the Appendix, changes
none of the physics relevant for this discussion.

The behavior of the 't Hooft loop of the B theory is familiar.
The potential energy at long distance between two ultraheavy 
spinors/monopoles is given by the weak coupling physics of
the A theory as $1/(r\log r)$, as for electrons in massless QED.

I will now argue that the spinor/monopole of finite mass
is well described as a monopole of the B theory when it is heavy,
and well described as a weakly coupled spinor of $Spin(8)$
when it is light.

Let us go along a flat direction $\vev{\det S}=v_0^{N_f-4}$ such
that the B theory breaks to $SO(N_f-4)$ at the scale
$v_0$.  If $v_0\gg\Lambda_B$, then this breaking occurs at weak
coupling, and the semiclassical monopole solution can be trusted.  In the A
theory, the mass of this state is much larger than $\Lambda_A$ 
(the scale of the perturbative Landau pole of the A theory) and so
the details of its structure are lost in a strongly coupled fog.
However, its long-range fields extend into the weakly coupled regime
of the A theory, and as they are unscreened, they must be the
electric fields of a massive spinor of $Spin(8)$.  Thus, from the
point of view of the low-energy $Spin(8)$ theory, the monopole of the
B theory will act as a ultramassive particle with quantum numbers of
an \ates\ representation.

As the scale $v_0$ is taken smaller and the monopole becomes lighter,
the B theory becomes strongly coupled and the semiclassical
description of the monopole will gradually worsen.  Meanwhile the
massive \ates\ particle of the A theory cannot decay, since it carries
a conserved $Z_2$ quantum number, and it must survive as a
light particle of the weakly coupled low-energy A theory.  General
renormalization group considerations ensure that, if it is light enough,
its properties will be those of an ordinary particle --- for example,
its kinetic terms will be canonical.  Where duality makes its
strongest statement is that this particle becomes massless as $v_0$
goes to zero.  Anomalies and other symmetry considerations make this
possible, and in a non-trivial way, necessary.

We therefore see that for large $N_f$ the heavy monopole description
gradually and smoothly goes over to the light spinor description as
$v_0$ is taken from large to small.  This uneventful transition
between two controllable regimes is possible because the B theory has
a free magnetic phase even when the spinor is massless, and so
reliable weakly coupled descriptions for the theory and its spinor/monopole 
exist both in the ultraviolet and in the infrared.

\subsec{The Non-Abelian Coulomb Phase}

Finally, consider $Spin(8)$ with $7\leq N_f\leq 17$ massless fields
$V^i$ in the \atev\ and a massive spinor $P$ in the \ates.  In this
case, the low-energy theory is a non-trivial conformal field theory.
The potential energy between spinor/monopole sources is $1/r$, which
follows from conformal invariance. 

No direct construction of these conformal field theories has been
found, and many of their properties have not been characterized.  As a
result, it is impossible to say much about the effect of the massive
spinor/monopole on the low-energy theory, other than to note its
presence will generate various irrelevant perturbations, whose form is
constrained by symmetries, on the low-energy fixed point.  Perhaps in
the future we will learn how to make more useful statements about this
physical situation.

\newsec{Final Remarks}

It would be enlightening to have many more examples of similar
phenomena.  An example of a theory which would be interesting to
understand is $Spin(8)$ with $N_f$ fields in the \ates\ and one in the
\atec\ or \atev, constructed by perturbing the $Spin(10)$ theory with 
multiple fields in the ${\bf 16}$ and ${\bf 10}$.  Although the A
theory is the same, up to $Spin(8)$ triality, as the $Spin(8)$ theory
studied in earlier sections, the B theory description will be very
different. It would be interesting to see if the \atev\ or \atec\ are
visible as monopoles in the dual theory, and if so, through what
symmetry breaking pattern.  Unfortunately, the B theory for $Spin(10)$
with multiple spinors is complicated by the presence of 
quantum accidental symmetries \refs{\NAD,\accid,\mltispnr}, 
and this construction has not yet been performed.

 However, there is in general no {\it a priori} reason to expect any
given \none\ supersymmetric duality transformation to map a massive
field to a topologically stable soliton.  The criteria under which a
particle of one theory will appear in the semiclassical physics of the
other have not yet been understood.  The absence of BPS charges for
particles means that there are usually no rigorous arguments.  It may
be hoped that future work in string theory will improve our
understanding of these issues.

The implications of the spinor-monopole identification for string
theory and M theory constructions of \none\ duality are worthy of
note.  The first chiral gauge theories from Type IIA string theory and
M theory branes are being constructed at the time of this writing
\chiralM.  The fact that spinors in $Spin(N)$ appear as non-BPS
monopoles of chiral gauge theories may be a useful hint for the Type
IIA or M theory construction of this duality.  Furthermore, it may
portend a substantial number of results to come involving non-BPS solitons in
field theory, string theory and M theory, along the lines of the
non-BPS strings found in the M theory construction of QCD \MQCD.  At the
very least, it will be interesting to understand how the relations
found in this paper are manifested in the Type IIA/M theory construction.

Finally, many \none\ gauge theories have yet to be fully understood.
The question of whether there are other unusual or perhaps
misidentified phases of these theories remains open.  The ability to
map Wilson loops into the dual description of a theory is critical for
testing its properties.  It may be hoped that further work in the
directions suggested by this paper will lead to greater insight into
the phase structure of gauge theories, and into the nature of duality
itself.

\centerline{Acknowledgements}

I thank my colleagues at the Institute for Advanced Study for advice
and interesting conversations.  I would especially like to acknowledge
conversations with J. Brodie, N. Seiberg, and E. Witten.  I would also
like to thank S.-J. Rey, S. Sethi and E. Weinberg for information
about monopoles, and K. Intriligator and N. Seiberg for reading the
manuscript.  This work was supported by the National Science
Foundation under Grant PHY-9513835 and by the WM Keck Foundation.

\appendix{A}{Duality Transformations for $Spin(N)$}

The five duality transformations used in this paper are
summarized.  Only essential elements are presented; the reader is
directed to the original references for more details.

\subsec{$Spin(N)$ with $N_f$ vectors}

The A theory has gauge group $SO(N)$; the B theory has
gauge group $SO(N_f-N+4)$ \NAD.  They share an $SU(N_f)$ global
symmetry.  The matter content of the theory is as follows:

\bigskip
\parasize=1in

\begintable
 A theory  \| $ SO(N)$ \| $[SU(N_f)] $ \| $$ 
\| B theory \| $SO(N_f-N+4) $ \| $ [SU(N_f)]$\crthick
$V^i $ \| $ {\bf N} $ \| $ \fund   $ \| \| $Q_i $ \| $\fund $ \| $ \antifund $
\nr
 \| $ \ $ \| $ \ $ \| $  $ \| $N^{ij}$ \| ${\bf 1}$ \| $\sym$ \endtable
\bigskip
%

Under duality the following chiral operators are identified:

\bigskip
\parasize=1in

\begintable
$V^iV^j $ \| $\leftrightarrow$ \| $N^{ij}$ \cr
$(V)^{N-2k}W_\alpha^k $ \| $\leftrightarrow$
   \| $(Q)^{N_f-N+2k}\tilde W_\alpha^{2-k}$ 
\endtable
\bigskip
%

The superpotential of the A theory is zero, while that
of the B theory, setting its coefficient to one, is 
\eqn\Weight{
W = N^{ij}Q_iQ_j}

\subsec{$Spin(8)$ with $N_f$ vectors and one spinor}

The A theory has gauge group $Spin(8)$; the B theory has
gauge group $SU(N_f-4)$ \ppmsone.  They share an $SU(N_f)$ global
symmetry.  The matter content of the theory is as follows:

\bigskip
\parasize=1in

\begintable
 A theory  \| $ Spin(8)$ \| $[SU(N_f)] $ \| $$ 
\| B theory \| $SU(N_f-4) $ \| $ [SU(N_f)]$\crthick
$V^i $ \| ${\bf  8_v} $ \| $ \fund   $ \| \| $S $ \| $\sym $ \| ${\bf 1}$\nr
$P $ \| $ {\bf 8_c} $ \| $ {\bf 1}$ \| $      $
 \| $Q_i $ \| $\antifund $ \| $ \antifund $
\nr
 \| $ \ $ \| $ \ $ \| $  $ \| $N^{ij}$ \| ${\bf 1}$ \| $\sym$ \nr
 \| $ \ $ \| $ \ $ \| $  $ \| $T$ \| ${\bf 1}$ \| ${\bf 1}$ \endtable
\bigskip
%

Under duality the following chiral operators are identified:

\bigskip
%
\parasize=1in

\begintable
$V^iV^j $ \| $\leftrightarrow$ \| $N^{ij}$ \cr
$PP $ \| $\leftrightarrow$ \| $T$ \cr
$(V)^4P^2 $ \| $\leftrightarrow$ \| $(Q)^{N_f-4}$ \endtable
\bigskip
%

The superpotential of the A theory is zero, while that
of the B theory, setting all coefficients to one, is 
\eqn\Weight{
W = N^{ij}Q_iSQ_j + T \det S}

\subsec{$Spin(8)$ with $N_f$ vectors and one spinor along with gauge singlets
\ppmsone}

This theory is a trivial modification of the previous one \ppmsone.
The matter content of the theory is as follows:

\bigskip
\parasize=1in

\begintable
 A theory  \| $ Spin(8)$ \| $[SU(N_f)] $ \| $$ 
\| B theory \| $SU(N_f-4) $ \| $ [SU(N_f)]$\crthick
$V^i $ \| ${\bf 8_v} $ \| $ \fund   $ \| \| $S $ \| $\sym $ \| ${\bf 1}$\nr
$P $ \| ${\bf  8_c} $ \| $ {\bf 1}$ \| $ $ 
    \| $Q_i $ \| $\antifund $ \| $ \antifund $ \nr
 $M_{ij}$ \| ${\bf 1}$ \| $\symbar$  \| $ \ $ \| $ \ $ \| $  $ \|\nr
 $U$ \| ${\bf 1}$ \| ${\bf 1}$  \| $ \ $ \| $ \ $ \| $  $ \|\endtable
\bigskip

%

Under duality the following chiral operators are identified:

\bigskip
%
\parasize=1in

\begintable
$M_{ij} $ \| $\leftrightarrow$ \| $Q_iSQ_j$ \cr
$U $ \| $\leftrightarrow$ \| $\det S$ \cr
$(V)^4P^2 $ \| $\leftrightarrow$ \| $(Q)^{N_f-4}$ \endtable
\bigskip
%

The superpotential of the B theory is zero, while that
of the A theory, setting all coefficients to one, is 
\eqn\Weight{
W = M_{ij}V^iV^j + UPP}

\subsec{$Spin(10)$ with $N_f$ vectors and one spinor}

The A theory has gauge group $Spin(10)$; the B theory has
gauge group $SU(N_f-5)$ \ppmstwo.  They share an $SU(N_f)$ global
symmetry.  The matter content of the theory is as follows:

\bigskip	
\parasize=1in

\begintable
   A  theory \| $ Spin(10)$ \| $[SU(N_f)] $ \| $ \ $ \|
B  theory \| $SU(N_f-5) $ \| $ [SU(N_f)]$\crthick
$V^i $ \| $ {\bf 10} $\| $ \fund   $ \| \| $S $ \| $\sym $ \| ${\bf 1}$\nr
$P $ \| $ {\bf 16} $ \| $ {\bf 1}$ \| $      $ 
     \| $Q_i $ \| $\antifund $ \| $ \antifund $
\nr 
\| $ \ $ \| $ \ $ \| $  $ \| $F$ \| $\fund$ \| ${\bf 1}$ \nr
 \| $ \ $ \| $ \ $ \| $  $ \| $N^{ij}$ \| ${\bf 1}$ \| $\sym$ \nr
 \| $ \ $ \| $ \ $ \| $  $ \| $Y^i$ \| ${\bf 1}$ \| $\fund$ \endtable
\bigskip

Under duality the following chiral operators are identified:

\bigskip
\parasize=1in

\begintable
$V^iV^j $ \| $\leftrightarrow$ \| $N^{ij}$ \cr
$V^iPP $ \| $\leftrightarrow$ \| $Y^i$ \cr
$(V)^5(P)^2 $ \| $\leftrightarrow$ \| $(Q)^{N_f-5}$ \endtable
\bigskip

The superpotential of the A theory is zero, while that
of the B theory, setting all coefficients to one, is 
\eqn\Wten{
W = \det S + N^{ij}Q_iSQ_j + Y^i Q_iF}

\subsec{$Spin(10)$ with $N_f$ vectors and $N_P>1$ spinors}

The A theory has gauge group $Spin(10)$; the B theory has gauge group
$SU(\tilde N)\times Sp(2\tilde M)$, where $\tilde N = N_f+2N_P-7$ and
$\tilde M=(N_P-1)$ \mltispnr.  They share an $SU(N_f)$ global
symmetry.  The $SU(N_P)$ symmetry which rotates the spinors is a
quantum accidental symmetry \accid\ in the B theory, which classically
has only an $SU(2)$ subgroup of this symmetry.  This $SU(2)$ is
embedded in $SU(N_P)$ such that the ${\bf N_P}$ representation of the
former is the ${\bf N_P}$ representation of the latter. For simplicity
I indicate the transformation properties of operators only under the
$SU(2)$ subgroup of $SU(N_P)$.

 The matter content of the theory is as follows:

\bigskip
\parasize=1in

\begintable
$ {\rm A \ theory}$ \| $ Spin(10)$ \| $[SU(N_f)] $ \| $[SU(2)]$
\|  \|$ {\rm B \ theory}$ \| $SU(\tilde N) $\| $Sp(2\tilde M) $ 
\| $ [SU(N_f)]$ \| $[SU(2)]$\crthick
$V^i $ \| $ {\bf 10} $ \| $ \fund   $ \| ${\bf 1}$\| \|
 $S $ \| $\sym $ \| ${\bf 1}$\| ${\bf 1}$\| ${\bf 1}$\nr
$P_I $ \| ${\bf  16} $ \| $ {\bf 1}$ \|$ N_P$ \|  \|
$Q_i $ \| $\antifund $ \|  ${\bf 1}$ \| $ \antifund $ \| ${\bf 1}$ \nr
 \| $ \ $ \| $ \ $ \| $  $ \| \| 
$F^r$ \| $\fund $ \| ${\bf 1}$ \|  ${\bf 1} $ \| ${\bf 2N_P-1}$
\nr 
\| $ \ $ \| $ \ $ \| $  $ \|\| 
$ Q'_a$ \| $\antifund $ \| $\fund$ \|  ${\bf 1} $ \| ${\bf 2}$
\nr 
\| $ \ $ \| $ \ $ \| $  $ \|\| 
$ t_X $ \| ${\bf 1}$ \| $\fund$ \|  ${\bf 1} $ \| ${\bf 2N_P-2}$
\nr 
 \| $ \ $ \| $ \ $ \| $ \ $ \| $  $ \| 
$N^{ij}$ \| ${\bf 1}$  \| ${\bf 1}$ \| $\sym$ \| ${\bf 1}$ \nr
 \| $ \ $ \| $ \ $ \| $ \ $ \| $  $ \| 
$Y^i_r$  \| ${\bf 1}$ \| ${\bf 1}$ \| $\fund$  \| $\bf{2N_P-1}$ \endtable
\bigskip

Under duality the following chiral operators are identified:

\bigskip
\parasize=1in

\begintable
$V^iV^j $ \| $\leftrightarrow$ \| $N^{ij}$ \cr
$V^i(P)^2 $ \| $\leftrightarrow$ \| $Y^i_a, (Q)^{N_f-1}(Q')^{2N_P-6}$ \cr
$(V)^3(P^2) $ \| $\leftrightarrow$ \| $(Q)^{N_f-3}(Q')^{2N_P-4}$ \cr
$(V)^5(P)^2 $ \| $\leftrightarrow$ \| $(Q)^{N_f-5}(Q')^{2N_P-2}$ \cr
$(P)^4 $ \| $\leftrightarrow$ \| $(t)^2,\dots$ \endtable
\bigskip
The superpotential of the A theory is zero, while that
of the B theory, setting all coefficients to one, is 
\eqn\Wtenmlti{
W =  N^{ij}Q_iSQ_j + Y^i_r Q_i F^r + Q'_a S Q'_b\epsilon^{ab}
+ Q'_a t_X F^r 
}

\listrefs
\end